\documentstyle[epsf,aps,multicol]{revtex}

\begin{document}

\def\thefootnote{\fnsymbol{footnote}} 
\title{Coexistence and Phase Separation in Sheared Complex Fluids}
\author{Peter~D. Olmsted$^{1\ast}$ and C.-Y.~D.~Lu$^{2\dagger}$}
\address{$^1$Department of Physics, University of Leeds, Leeds LS2 9JT,
UK and $^2$Polymer and Colloids, Cavendish Laboratory, Cambridge
University, Madingley Road, Cambridge CB3 0HE, UK} 
\date{\today} 
\maketitle

\begin{abstract} 
We demonstrate how to construct dynamic phase diagrams for complex fluids
that undergo transitions under flow, in which the conserved composition
variable and the broken-symmetry order parameter (nematic, smectic,
crystalline, {\sl etc.\/} ) are coupled to shear rate. Our construction
relies on a selection criterion, the existence of a steady interface
connecting two stable homogeneous states. We use the (generalized) Doi
model of lyotropic nematic liquid crystals as a model system, but the
method can be easily applied to other systems, provided non-local effects
are included.
\end{abstract}

\begin{multicols}{2}

Complex fluids in shear flow display a range of
behaviors which is only beginning to be unearthed
\cite{safinya,berret,rehage,diat,schmitt94,callaghan96}.  Shear can
perturb equilibrium phase transitions ({\sl e.g.} the isotropic-to-nematic
(I-N) liquid crystalline \cite{berret,hess76,olmsted,see90} and
isotropic-to-lamellar \cite{catesmilner} transitions), and induce
structures, ({\sl e.g.\/} bilayer onions \cite{diat}) that exist only as
metastable equilibrium phases. A related phenomenon is dynamic instability
in non-Newtonian fluids, when the theoretical {\sl homogeneous\/}
stress--strain-rate constitutive relation exhibits multi-valued behavior,
as in theories of polymer melts \cite{doiedwards,catesmcleish93} and
worm-like micelles \cite{spenley93,cates90}. Such models are often
used to describe, for example, the spurt effect, whereby the flow
rate of a fluid in a pipe changes discontinuously as a function
of applied pressure drop \cite{mcleish86}.  The most important
unresolved question about non-monotonic flow curves, such as those in
Fig.~\ref{fig:both}a, is: what determines the stress, if any, at which
the system phase separates into `bands'? Suggestions have included
(a) variational hypotheses (Ref.~\cite{porteunpub}); (b) assuming the
stress at the top of the stable viscous branch is selected (``top jump")
\cite{catesmcleish93,spenley93,schmitt95}; (c) including geometrical
effects \cite{greco96}; and (d) incorporating (physically present)
non-local contributions to the stress \cite{olmsted,pearson94,spenley96}.
In this Letter we pursue (d) and explore in detail the utility of
constitutive inhomogeneous effects in resolving the issue of stress
selection in complex fluids, which has occupied the rheology and physics
communities in the guise of either unstable flows \cite{pearson94} or
non-equilibrium phase transitions \cite{olmsted}.

After a general discussion we introduce semi-phenom\-enological equations
of motion for rod-like molecules in solution, extending the Doi model
\cite{doi} to inhomogeneous flows.  Our study has the following goals
and results: (1) We present a {\sl general\/} recipe for computing
phase separation under flow, and hence the experimentally measured
rheological behavior; (2) We point out that the proper field variable
({\sl either\/} stress {\sl or\/} strain rate) may not be unique, a
feature absent from equilibrium systems; (3) Using concepts developed
for dynamical systems theory we conclude that stress selection of models
with non-local (in space) differential constitutive relations in planar
shear flow is {\sl unique\/}; {\sl i.e.\/}, as with equilibrium phase
transitions, it occurs along a hyper-surface of lower dimension than that
of the field variable space.  Our discussion is facilitated by examining
the stress--strain-rate--composition surface, a representation we have
not seen before and hope will become commonplace.

{\begin{figure}
\par\columnwidth20.5pc
\hsize\columnwidth\global\linewidth\columnwidth
\displaywidth\columnwidth
\epsfxsize=3.5truein
\centerline{\epsfbox{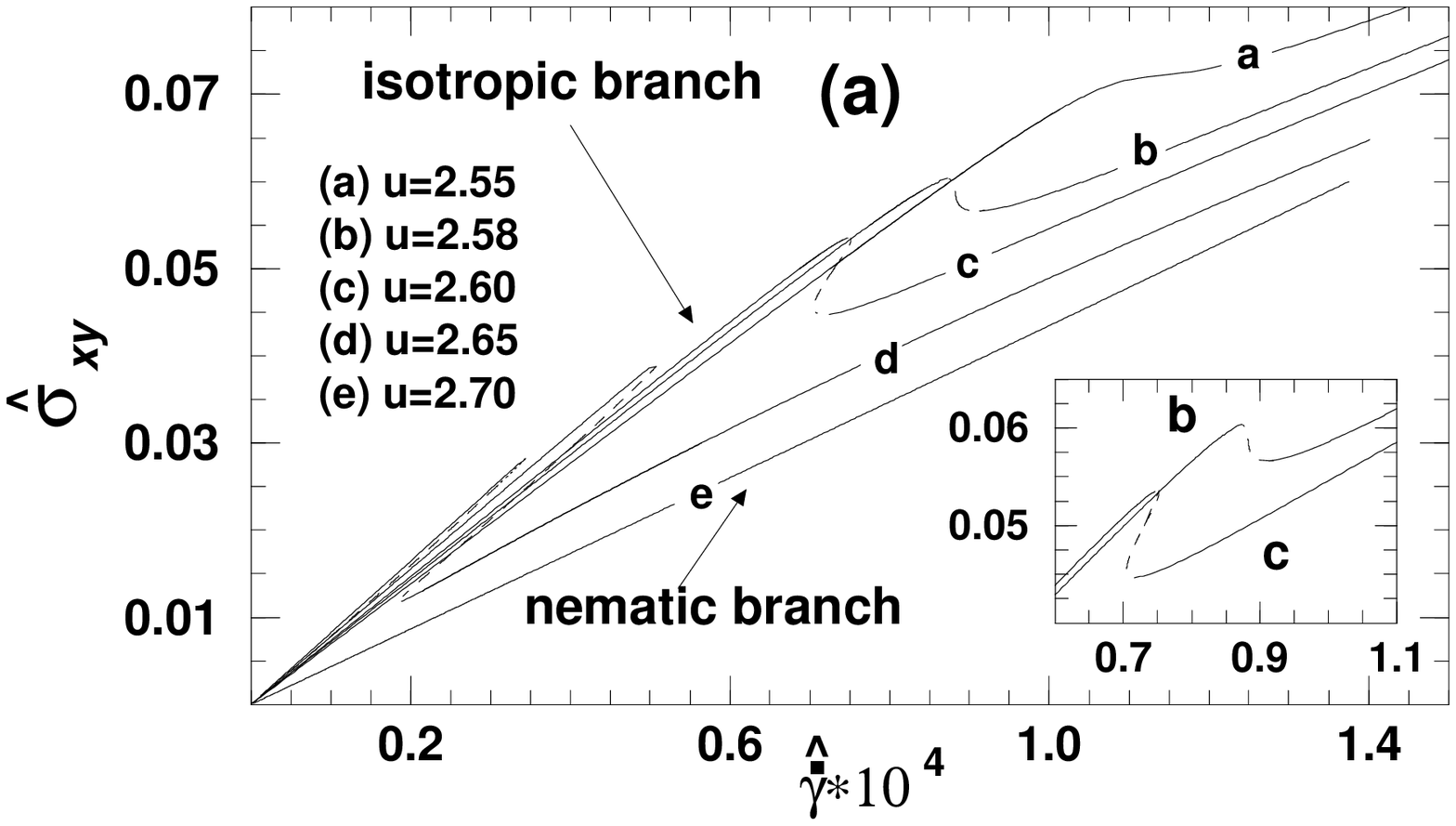}}
\epsfxsize=3.5truein
\centerline{\epsfbox{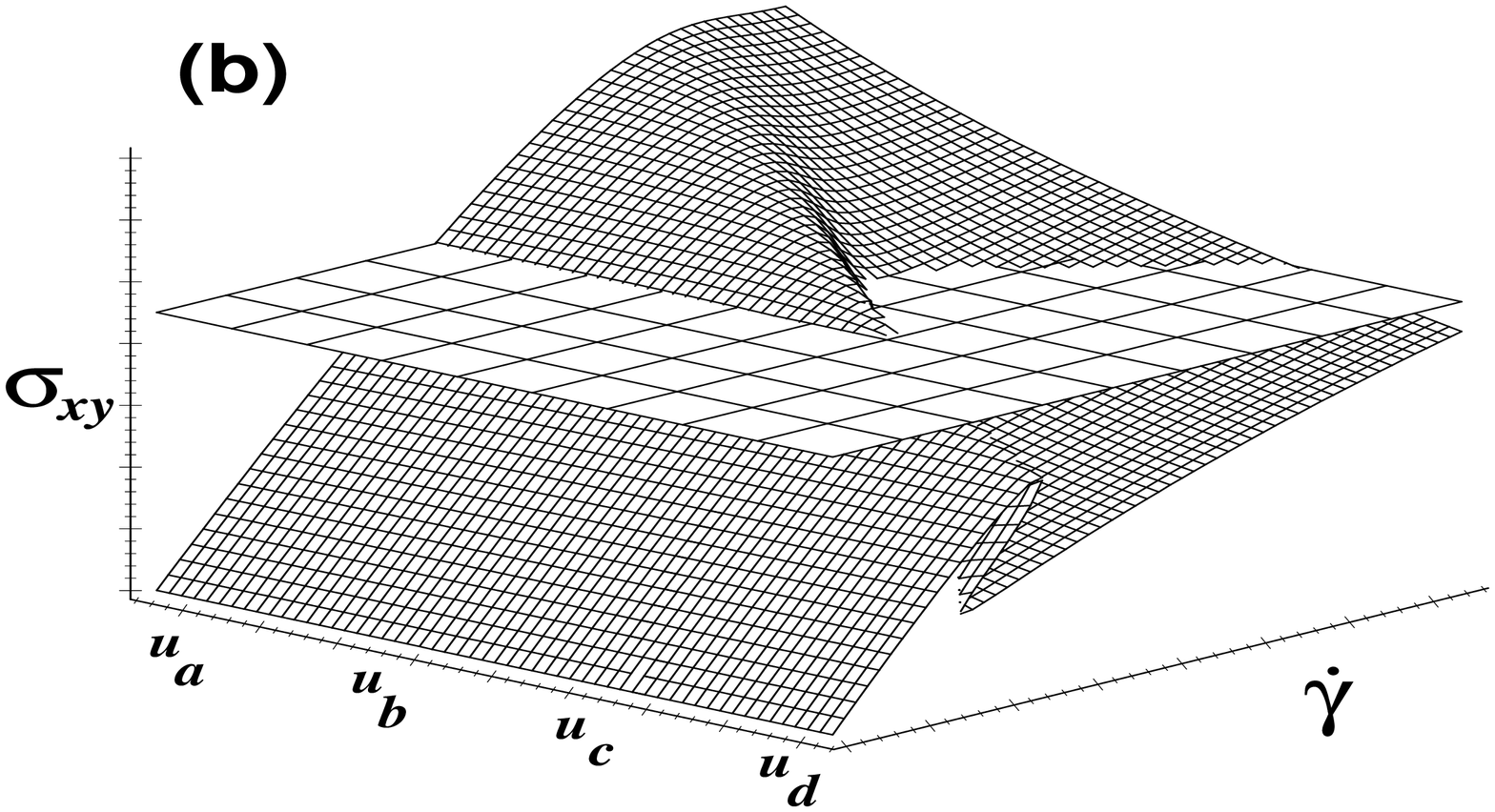}}
\caption{(a) Homogeneous stress $\hat{\sigma}_{xy}$ vs. strain rate
$\widehat{\dot{\gamma}}$ behavior for various excluded volumes
$(u\equiv\phi L \nu_2/\alpha)$ as calculated from the Doi model
with quadratic closure [21,23].  We use the dimensionless parameters
$\widehat{\dot{\gamma}}\equiv\dot{\gamma} /(6 D_{ro}\nu_1\nu_2^2)$
and $\hat{\sigma}_{xy}\equiv \sigma_{xy} \nu_2\log L / (9\pi\eta_s
D_{ro})$. Dotted lines mark unstable branches.  The plane in (b) is at
fixed stress. At $u_e$ both isotropic and nematic phases are metastable
in zero shear. Here and below $L=5$.}
\label{fig:both}
\end{figure}}
The curves in Fig.~\ref{fig:both}a are reminiscent of pressure-volume
isotherms in a liquid-gas system and, accordingly, we seek to construct a
`phase diagram' by pursuing an analogy between homogeneous stable steady
states and equilibrium phases.  As in equilibrium, non-equilibrium
`phases' may be separated, in field variable space, by hyper-surfaces
representing continuous ({\sl e.g.\/} critical points)  or discontinuous
(`first-order') transitions \cite{olmsted,safinya}. Coexistence implies
an inhomogeneous state spanning separate branches of the homogeneous flow
curves. In contrast to equilibrium bulk phases, where field variables are
uniquely identified, here we may consider phase coexistence at a given
shear stress, with the interface parallel to the vorticity-velocity plane
(curves $b\!-\!e$); {\sl or\/} a given strain rate (curves $c\!-\!e$),
with the interface parallel to the velocity-velocity gradient plane. The
appropriate field variable is thus determined by the nature of the
constitutive relation (for curve $b$, common strain rate is impossible),
or even by the flow history or the rheometer ({\sl e.g.\/} fixing the
stress or strain rate in curves $c\!-\!e$). In this Letter we compute
coexistence at a common stress, but note that computation for a common
strain rate is analogous.  In equilibrium, minimizing the total
free energy leads to equality of field variables between two phases and
the Maxwell common tangent condition ({\sl e.g.\/} the equal osmotic
pressure condition, aided by equal chemical potential, in rod suspensions
\cite{lekkerkerker}). In dynamics, the former follows from the equations
of motion plus the stationary condition, with an extra shear degree of
freedom. However, the lack of a criterion to replace the latter leaves
an unresolved degeneracy ({\sl e.g.\/} which stress is selected for a
given averaged strain rate). Flow experiments on worm-like micelles
find a well-defined transition stress for the onset of banded flows
\cite{berret,rehage,schmitt94,callaghan96}, which suggests that the
degeneracy is not physical.

For the thermotropic ({\sl i.e.\/} restricted to the melt composition)
I-N transition under flow, Olmsted and Goldbart \cite{olmsted} resolved
this degeneracy by rejecting those possibilities which did not admit a
stationary interface solution and found, numerically, an apparently
uniquely selected state. They included {\sl gradient terms\/},
which penalize (energetically) spatial microstructure variations and
dictate the interface structure. The importance of gradient terms was
also recognized in Refs.~\cite{elkareh89,pearson94,spenley96}. The
existence condition of a stationary interface selects one among a band
of possible coexisting solutions, and is fundamentally different from
augmenting a local constitutive model with a variational principle
\cite{porteunpub} or assuming selection at the limits of stability
\cite{catesmcleish93,spenley93,schmitt95}. Below we justify the
{\sl uniqueness\/} of a selected state. 

Now we proceed with our calculation.
The free energy of a solution of rod-like molecules of volume fraction
$\phi$ in an athermal solvent 
({\sl e.g.\/} as in Ref.~\cite{lekkerkerker}) is given by 
\begin{eqnarray}
{\cal F\/}(\phi,\bbox{Q})&& =k_{\scriptscriptstyle B}T \int\!d^3\!r
\left\{\phantom{\bigl\{}\!\!\!{\phi\over v_r}\log\phi +
{\left(1\!-\!\phi\right)\over v_s}\log\left(1\!-\!\phi\right) 
\right. \nonumber\\ 
&&
+ {\phi\over v_r} 
\left[ \case12 \left(1\!-\!\case13 u\right)\hbox{\rm Tr}
\bbox{Q}^2\!
-\case13 u \hbox{\rm Tr}\bbox{Q}^3\!
+\case14 u\left(\hbox{\rm Tr}\bbox{Q}^2\right)^2 
\right.
\nonumber\\
&& 
\left.\left. + \case12 K\left(\nabla_{\alpha}Q_{\beta\lambda}\right)^2 \right]
 + \case12 g \left(\nabla\phi\right)^2 \right\}.
\label{eq:free}
\end{eqnarray}
Here $v_r$ and $v_s$ are rod and
solvent monomer volumes;
 $Q_{\alpha\beta}$ is the nematic order parameter tensor
\cite{pgdg}; and $u\equiv\phi L \nu_2/\alpha$  is Doi's excluded volume
parameter \cite{doi}, where $L$ is the rod aspect ratio and $\nu_2$ and
$\alpha$ are ${\cal O\/}(1)$ geometrical prefactors. This free energy
includes the entropy of mixing, the orientational contribution
yielding an I-N transition \cite{doi}, and 
spatial correlations of composition and nematic order within the
one constant approximation. The phenomenological gradient terms 
may, in principle, be calculated from a microscopic model.

The equations of motion are \cite{doi,tobepub,hohenberg77}:
\begin{eqnarray}
\rho\left(\partial_t + {\rm\bf v}\cdot\bbox{\nabla}\right) {\rm\bf v} &=&
\bbox{\nabla}\!\cdot\!\bbox{\sigma} (\phi, \bbox{\kappa},\bbox{Q}) 
\label{eq:1} \\ 
\left(\partial_t + {\rm\bf v}\cdot\bbox{\nabla}\right) \bbox{Q} &=& 
\bbox{F}(\bbox{\kappa},\bbox{Q}) + \bbox{G} (\phi, \bbox{Q}) 
\label{eq:2} \\
\left(\partial_t + {\rm\bf v}\cdot\bbox{\nabla}\right) \phi &=& -\nabla
\cdot{\rm\bf J}, \label{eq:3}
\end{eqnarray}
with $\kappa_{\alpha\beta} = \nabla_{\beta} v_{\alpha}$ and $\rho$ the
density. The stress tensor is
\begin{equation}
\bbox{\sigma} = -p{\rm\bf I} + 2\eta \bbox{\kappa}^{s} + 
\bbox{\sigma}_{\mit rev}(\phi,
\bbox{Q}),
\end{equation}
where 
$\kappa^s_{\alpha\beta}\equiv(\kappa_{\alpha\beta} +
\kappa_{\beta\alpha})/2$
and we take $\eta$ to be the solvent viscosity $\eta_s$ for simplicity.
The reversible stress due to the nematic 
order is \cite{doi,olmsted,tobepub}
\begin{equation}
\bbox{\sigma}_{\mit rev} = - 3\bbox{H} + 
\bbox{H}\!\cdot\!\bbox{Q} - \bbox{Q}\!\cdot\!\bbox{H} 
- \bbox{\nabla}Q_{\alpha\beta}
{\delta{\cal F\/}\over\delta\bbox{\nabla}Q_{\alpha\beta}},
\end{equation}
where $\bbox{H} = -\delta{\cal F\/}/\delta \bbox{Q}$. The isotropic stress 
from the force density $(\bbox{\nabla}\!\phi) \delta
{\cal F\/}/\delta\phi$ \cite{hohenberg77} has been neglected.
In Eq.~(\ref{eq:2}) the (reactive) ordering term $\bbox{F}$
is given by
\begin{equation}
\bbox{F}(\bbox{\kappa},\bbox{Q})\!=\!\case23\bbox{\kappa}^{s}\!+ 
\!\bbox{\kappa}\!\cdot\!\bbox{Q}\!+\!
\bbox{Q}\!\cdot\!\bbox{\kappa}^{\scriptscriptstyle T}
\!-\!2(\bbox{Q}\!+\!\case13\bbox{I})\hbox{\rm
Tr}(\bbox{Q}\!\cdot\!\bbox{\kappa}).
\end{equation}
For simplicity, we have chosen the form appropriate for an infinite aspect 
ratio \cite{doi}. 
The dissipative portion $\bbox{G}$ is 
\begin{equation} 
\bbox{G}(\phi,\bbox{Q}) = {6 \nu_1 D_{\mit r0} v_r^3\over
k_{\scriptscriptstyle B}T \phi^3 \ell^6 
(1-\case32\hbox{\rm Tr}\bbox{Q}^2)^2} \bbox{H},
\end{equation} 
where $D_{\mit r0}$ is the single-rod rotational diffusion coefficient,
$\nu_1$ is a geometric prefactor, and $\ell$ is the rod length \cite{doi}.
The chemical potential $\mu$ drives the current ${\rm\bf J}$, 
\begin{equation}
{\rm\bf J} = - \bbox{M} \cdot \bbox{\nabla} \mu,
\end{equation}
where $\bbox{M}$ is the mobility tensor and
$\mu \equiv \delta {\cal F\/}/\delta\phi$.

For other systems, equations like Eqs.~(\ref{eq:3}) and (\ref{eq:2})
govern the conserved and broken-symmetry (or other long-lived) variables,
respectively. For some local models, internal dynamics (Eq.~\ref{eq:2})
can be eliminated to give the stress as a history integral over
the strain rate. In polymer melts \cite{doiedwards}, and worm-like micelles
\cite{cates90} far from a nematic regime, this leads to non-monotonic
stress--strain-rate curves.

We seek stable steady-state solutions to Eqs.~(\ref{eq:1}-\ref{eq:3}) for
planar shear, ${\rm\bf v}({\rm\bf r}) = \dot{\gamma}y{\bf\hat{x}}$.
Integrating Eqs.~(\ref{eq:1},\ref{eq:3}) along $y$ yields
$\sigma_{xy}(\phi, \dot \gamma, \bbox{Q})=\sigma_0$ and $\mu(\phi,
\bbox{Q})=\mu_0$, where $\sigma_0$ is the applied stress.  One integration
constant of Eq.~(\ref{eq:3}) is zero from the boundary condition
$J_y\!=\!0$, while $\mu_0$ is determined below \cite{pathology}.
For homogeneous solutions, $\bbox{Q}$ may be eliminated from
$\sigma_{xy}$ and $\mu$ using Eq.~(\ref{eq:2}). The stress is shown
in Fig.~\ref{fig:both}. Because ${\cal F\/}(\phi,\bbox{Q})$ describes
an I-N transition, multiple roots for $\bbox{Q}$ may exist at a given
stress \cite{olmsted,see90}, with distinct strain rates.

{\begin{figure}
\par\columnwidth20.5pc
\hsize\columnwidth\global\linewidth\columnwidth
\displaywidth\columnwidth
\epsfxsize=3.25truein
\centerline{\epsfbox{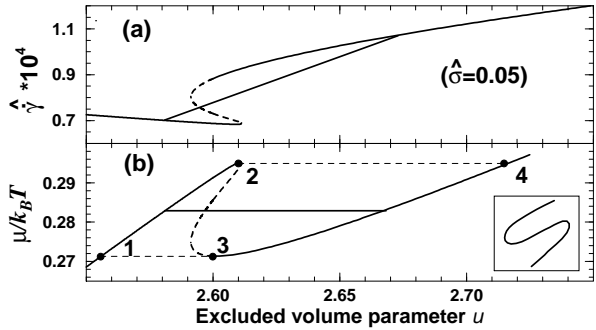}}
\caption{(a) Reduced strain rate $\widehat{\dot{\gamma}}(u)$ for the
stress contour in Fig.~\protect{\ref{fig:both}b}. 
(b) $\mu(u)$ along the curve 
in (a). A selected pair has been
indicated in (a) and (b). The inset shows another (topologically) possible 
$\mu(u)$. Dashed lines mark the limits of stability
of coexisting pairs.} 
\label{fig:strainmu}
\end{figure}}
For coexistence at fixed stress $\sigma_0$ (the interface has a
normal parallel to $\hat{y}$), a stress contour traces a line in the
$(\dot{\gamma}\!-\!\phi)$ plane (Fig~\ref{fig:strainmu}a). We plot
$\mu(\phi)$ along this line in Fig.~\ref{fig:strainmu}b.  Coexisting
phases must have the same $\mu$.  For concentrations $\bar{\phi}$
outside points 1 and 4, $\mu_0$ is determined uniquely.  Between these
points, we must determine $\mu_0$ at which two phases coexist. Following
Ref.~\cite{olmsted}, we select the $\mu_0$ which permits a stationary
inhomogeneous solution satisfying $\sigma_{xy}\!=\!\sigma_0$,
$\mu\!=\!\mu_0$ and Eq.(\ref{eq:2}) ($0=\bbox{F}+\bbox{G}$).  {\sl This
is not an auxiliary assumption, but follows from the inhomogeneous
equations of motion.\/} Given a selected $\mu_0$ and a mean concentration
$\bar\phi$, the portions of the two phases are fixed by the lever
rule, and the mean shear rate can be calculated.  In the static
limit ($\dot \gamma=0$) the stationary interface condition 
follows from minimizing ${\cal F}(\phi,
\bbox{Q})$ with a given $\bar \phi$ constraint. Functional minimization
with respect to $\phi(y)$ allows the interface position to move, thereby
adjusting the ratio of the two phases and recovering the common tangent
construction.

To see how $\mu_0$ can be selected uniquely, note that the stationary
solutions to Eq.(\ref{eq:2}), $\sigma_{xy}\!=\!\sigma_0$, and
$\mu\!=\!\mu_0$, are nonlinear ordinary differential equations
(ODE) involving $\partial / \partial y$.  These ODE may be converted
into an equivalent set of first order ODE with dependent variables
$\{\phi, d\phi/dy, \dot{\gamma}, \bbox{Q}, d\bbox{Q}/dy\}$.  In the ODE
phase space, the interfacial solution corresponds to a trajectory
(a `heteroclinic saddle connection') joining two fixed points (the
homogeneous states). As $\mu_0$ changes for fixed $\sigma_0$, the
phase flow changes catastrophically at those isolated values of $\mu_0$
where the desired trajectory exists.  [This is proven for differential
constitutive relations in planar flow \cite{tobepub} by showing that the
saddle connection, assuming it exists, is of the non-transverse type
\cite{saddle}].  This explains the apparent uniqueness of a selected
$\mu_0$, given $\sigma_0$, found numerically. Our (one-dimensional)
solution supports the {\sl existence\/} of such a solution in the modified
Doi model, and is stable against perturbations in $y$. Because non-local
effects (see Ref.~\cite{elkareh89,pearson94} for diffusion effects)
always exist in reality, and pathological degeneracies only occur in
local models, we expect that models that can resolve the interface
structure have unambiguous phase diagrams.

{\begin{figure}
\par\columnwidth20.5pc
\hsize\columnwidth\global\linewidth\columnwidth
\displaywidth\columnwidth
\epsfxsize=3.0truein
\centerline{\epsfbox{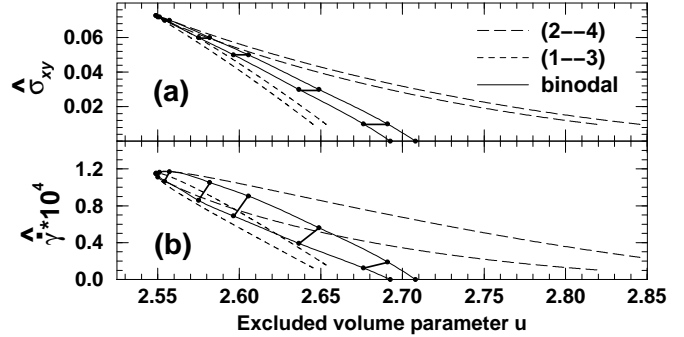}}
\caption{
Phase diagram in the $(\hat{\sigma}_{xy}\!-\!u)$ (a) and 
$(\widehat{\dot{\gamma}}\!-\!u)$ (b) planes.
Broken lines trace the loci of points
$1\!-\!4$ in Fig.~\protect{\ref{fig:strainmu}}, and the solid
lines are calculated tie lines.}
\label{fig:tie}
\end{figure}}
The selection criterion determines the tie lines on
Fig.~\ref{fig:strainmu}, and varying the stress yields the phase diagram of
Fig.~\ref{fig:tie}b \cite{tobepub}. In the latter, for $\dot{\gamma}=0$
the tie line is horizontal, and for $\dot{\gamma}>0$ tie lines have
positive slope, because the more concentrated nematic phase has a lower
effective viscosity.  For models where $\mu(\phi)$ has the shape
of the inset of Fig.~\ref{fig:strainmu}b, tie lines have negative slopes.

{\begin{figure}
\par\columnwidth20.5pc
\hsize\columnwidth\global\linewidth\columnwidth
\displaywidth\columnwidth
\epsfxsize=3.5truein
\centerline{\epsfbox{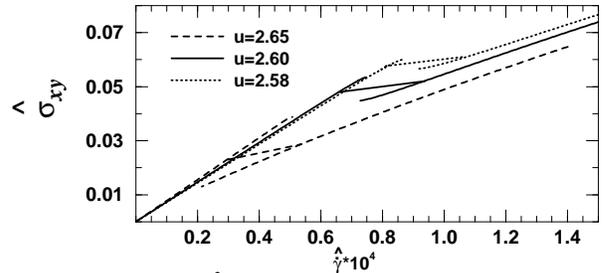}}
\caption{$\hat{\sigma}_{xy}$ vs. $\widehat{\bar{\dot{\gamma}}}$ 
for various compositions. 
To read this from Fig.~\protect{\ref{fig:tie}}, choose a $u$ and increase 
the stress. In the two-phase region jump from stress
tie line to stress tie line until the one phase region is reached.}
\label{fig:stress-strainbar}
\end{figure}}
Fig.~\ref{fig:stress-strainbar} shows the stress--averaged-strain-rate
curves as would be measured in an experiment. Here the coexistence
region comprises lines with positive slopes. For models with vertical
tie lines in Fig.~\ref{fig:tie}b, indicating a composition-independent
transition (as in the Doi-Edwards model of polymer melts), the plateau
would be flat. This is a graphical expression of the explanation of
a sloped plateau given by Schmitt {\sl et al.\/} \cite{schmitt95}
(however, they assumed ``top-jumping"). In general one must vary both
average concentration and strain rate to maintain constant shear stress, a
situation familiar from equilibrium multi-component phase coexistence. The
`plateau' need not be a straight line; its shape is determined
by the change in splay of the tie lines with increasing strain rate in
Fig.~\ref{fig:tie}b. Experiments on wormlike micelles \cite{berret}
have found the plateaus to be steady states and the `spines' to be
metastable branches.

In summary, the steps to compute phase separation under flow are:
(1) Determine the full {\sl inhomogeneous\/} equations of motion;
(2) Analyze stable homogeneous flows. (3) Choose the candidate field
variable for phase coexistence.  Multiple choices ({\sl e.g.\/}
stress or strain rate, according to interface geometry) must be
analyzed separately.  (4) Identify pairs of phases with the same field
variables, as in Fig.~\ref{fig:both}b and Fig.~\ref{fig:strainmu}b for
$\sigma_{xy}$ and $\mu$. (5) Determine the phase diagram (in field
variables) by requiring a stationary interface between homogeneous
solutions. To compare with experiments, density variables can be used,
as in Fig.~\ref{fig:tie}, to find (6) the tie lines where the lever
rule applies and (7) the space-averaged stress--strain-rate relations
(Fig.~\ref{fig:stress-strainbar}).  Other multi-component systems retain
the structure of Eqs.~(\ref{eq:1}-\ref{eq:3}), with Eq.~(\ref{eq:2})
governing the dynamics of some structural order parameter.

We have calculated coexistence at fixed stress, while curves
$c\!-\!e$ in Fig.~\ref{fig:both}a admit, in principle, coexistence at
fixed strain rate. A full solution of Eqs.~(\ref{eq:1}-\ref{eq:3})
requires analyzing both interface directions, which is reminiscent
of, and may be relevant to, the problem of the orientation of
diblock lamellar phases in shear \cite{goulian95}. Also, we have only
considered flow-aligning solutions to the Doi model. Another, so-called
`log-rolling', solution is stable at high shear rates \cite{bhave93} and
will be included in a complete treatment \cite{tobepub}.  The extended
Doi model is appropriate for hard rod suspensions \cite{lekkerkerker},
and we welcome experiments on these systems.  The flow instability is due
to perturbation of an equilibrium phase transition; while systems such
as worm-like micelles probably possess some combination of perturbed
(I-N) and dynamic transitions \cite{schmitt94,cates90}, which could
yield a stress--strain-rate--composition surface (Fig.~\ref{fig:both}b) with
multiple folds. 

We have not considered the important problem of the stability of an
undulating interface, which may restrict the choice of field variables
\cite{goulian95}. Stability analysis by Renardy \cite{renardy} on banded
flow in the Johnson-Segelmann model (which has the qualitative behavior
of curve $b$ in Fig.~\ref{fig:both}a, but no gradient terms) revealed a
stationary interface at {\sl any} stress in the two-stress region, and
a family of unstable high-wavenumber undulations. Gradient terms break
the stress degeneracy \cite{pearson94,spenley96,tobepub} and introduce a
stabilizing surface tension and dissipation within the interface as it
moves, which affects the stability analysis. Finally, complementary to
the planar shear case analyzed above, Greco and Ball \cite{greco96}, have
recently demonstrated the important result that, for a Johnson-Segelmann
fluid in {\sl cylindrical\/} Couette flow, a stationary interface exists
at a selected stress: coexistence is apparently influenced by the curved
boundary geometry of the flow.

We thank J.-F. Berret, M. Cates, F. Greco, P. Goldbart,
J. Harden, R. Larson, G. Leal, T. McLeish, G. Porte, and N. Spenley for
many discussions; the Isaac Newton Institute, where this work was begun;
and (C.-Y.~D.~L.) the Colloid Technology Programme for support.

\end{multicols}

\begin{thebibliography}{10}
\bibitem[\ast]{pdo} e-mail: phy6pdo@irc.leeds.ac.uk

\bibitem[\dagger]{dlu} e-mail: cydl1@phy.cam.ac.uk

\bibitem{safinya}
C.~R. Safinya, E.~B. Sirota, and R.~J. Plano, Phys. Rev. Lett. {\bf 66}, 
1986 (1991).

\bibitem{berret}
J.-F. Berret, {\it el al.\/}, Europhys. Lett. {\bf 25},
 521 (1994); J.-F. Berret, D.~C. Roux, and G. Porte, J.~Phys.~II (France) {\bf 4}, 1261 (1994).

\bibitem{rehage}
H.~Rehage and H.~Hoffmann, Mol. Phys. {\bf 74}, 933 (1991).

\bibitem{diat}
D. Roux, F. Nallet, and O. Diat, Europhys. Lett. {\bf 24}, 53 (1993).

\bibitem{schmitt94}
V. Schmitt, {\it et al.\/}, Langmuir {\bf 10}, 955 (1994).

\bibitem{callaghan96}
P.~T. Callaghan, {\it et al.\/}, J.~Phys.~II (France) {\bf 6}, 375 (1996).

\bibitem{hess76}
S.~Hess, Z.~Naturforsch. {\bf 31a}, 1507 (1976).

\bibitem{olmsted}
P.~D. Olmsted and P.~M. Goldbart, Phys.~Rev. {\bf A41}, 4578 (1990);
{\sl ibid\/}, {\bf A46}, 4966 (1992).

\bibitem{see90}
H. See, M. Doi, and R. Larson, J.~Chem. Phys. {\bf 92}, 792 (1990).

\bibitem{catesmilner}
M.~E.~Cates and S.~T. Milner, Phys. Rev. Lett. {\bf 62}, 1856 (1989).

\bibitem{doiedwards}
M. Doi and S.~F. Edwards, {\em The Theory of Polymer Dynamics} (Clarendon,
Oxford, 1989).

\bibitem{catesmcleish93}
M.~E. Cates, T.~C.~B. McLeish, and G. Marrucci, Europhys. Lett. {\bf 21}, 451
 (1993).

\bibitem{spenley93}
N.~A. Spenley, M.~E. Cates and T.~C.~B. McLeish, Phys. Rev. Lett. {\bf
71}, 939 (1993). 

\bibitem{cates90}
M.~E. Cates, J.~Phys. Chem. {\bf 94}, 371 (1990).

\bibitem{mcleish86}
T.~C.~B. McLeish and R.~C. Ball, J.~Poly. Sci. B-Poly. Phys. {\bf 24}, 1735
 (1986).

\bibitem{porteunpub}
G.~Porte, J.-F.~Berret, and J.~L.~Harden, unpublished (1996).

\bibitem{schmitt95}
V. Schmitt, C.~M. Marques, and F. Lequeux, Phys.~Rev. {\bf E52}, 4009 (1995).

\bibitem{greco96} F.~Greco and R.~C.~Ball, to be published (1996).

\bibitem{pearson94}
J.~R.~A. Pearson, J.~Rheol. {\bf 38}, 309 (1994).

\bibitem{spenley96}
N.~A. Spenley, X.~F. Yuan, and M.~E. Cates, J.~Phys.~II (France) {\bf 6}, 551
(1996).

\bibitem{doi}
M. Doi, J.~Poly. Sci: Poly. Phys. {\bf 19}, 229 (1981);
N. Kuzuu and M. Doi, J.~Phys. Soc. Jap. {\bf 52}, 3486 (1983).

\bibitem{lekkerkerker} P.~A.~Buinin and H.~N.~W.~Lekkerkerker, 
J.~Phys. Chem. {\bf 97}, 11510 (1993).

\bibitem{elkareh89}
A.~W.~El-Kareh and L.~G.~Leal, J. Non-Newt. Fl.~Mech. {\bf 33}, 257 (1989).

\bibitem{tobepub}
P.~D.~Olmsted and C.-Y.~D.~Lu, to be published.

\bibitem{pgdg}
P.~G. de~Gennes and J. Prost, {\em The Physics of Liquid Crystals}, 2nd ed.
 (Clarendon, Oxford, 1993).

\bibitem{hohenberg77}
P.~C. Hohenberg and B.~I. Halperin, Rev. Mod. Phys. {\bf 49}, 435 (1977).

\bibitem{pathology} In the few cases where the final phase diagram of
$\sigma_{xy}$ and $\mu$ has a transition line parallel to the $\mu$
axis, one must first fix $\mu_0$, and then determine $\sigma_0$.

\bibitem{saddle} See, {\it e.g.}, R.~H.~Abraham and C.~D.~Shaw, {\em
Dynamics - The Geometry of Behavior}, Part 3 (Aerial Press, Santa Cruz,
1985), p.53.

\bibitem{bhave93}
A.~V. Bhave, {\it et al.\/}, J.~Rheol. {\bf 37}, 413 (1993).

\bibitem{goulian95}
M.~Goulian and S.~T.~Milner, Phys. Rev. Lett. {\bf 74}, 1775 (1995).

\bibitem{renardy}
Y.~Y. Renardy, Th.~Comp. Fl.~Mech. {\bf 7}, 463 (1995).

\end{thebibliography}
\end{document}